\documentclass[]{aa}
\usepackage{natbib}
\bibpunct{(}{)}{;}{a}{}{,}
\usepackage{subfig}
\usepackage{graphicx}
\usepackage{epsf}
\usepackage{rotating}
\usepackage{graphics}
\usepackage{txfonts}
\newcommand{\corot}{\emph{CoRoT}}
\newcommand{\flames}{\emph{FLAMES}}
\newcommand{\giraffe}{\emph{GIRAFFE}}
\newcommand{\uves}{\emph{UVES/FLAMES}}
\newcommand{\MJ}{M$_{Jup}$}
\newcommand{\kms}{km\,s$^{-1}$\/}
\newcommand{\ms}{m\,s$^{-1}$\/}
\begin{document}
   \title{Doppler search for exoplanet candidates and binary stars in a \corot\/ field using a
     multi-fiber spectrograph \thanks{Based on observations collected with the
       \giraffe\ and \uves\ spectrographs at the VLT/UT2 Kueyen telescope (Paranal
       observatory,ESO,Chile: program 074.C-0633A).}}

   \subtitle{I. Global analysis and first results}

   \author{B. Loeillet\inst{1}
      \and
      F. Bouchy\inst{1,2}
      \and
      M. Deleuil\inst{1}
      \and
      F. Royer\inst{3}
      \and
      J.C. Bouret\inst{1}
      \and
      C. Moutou\inst{1}
      \and
      P. Barge\inst{1}
      \and
      P. de Laverny\inst{4}
      \and
      F. Pont\inst{5}
      \and
      A. Recio-Blanco\inst{4}
      \and
      N. C. Santos\inst{5,6}
      }

   \offprints{B. Loeillet,
     \email benoit.loeillet@oamp.fr}

   \institute{Laboratoire d'Astrophysique de Marseille, BP 8, 13376 Marseille
     cedex 12, France, Universit\'e de Provence, CNRS (UMR 6110) and CNES
          \email{benoit.loeillet@oamp.fr}
     \and
         Institut d'Astrophysique de Paris, CNRS, Universite Pierre et Marie Curie, 98$^{bis}$ Bd Arago, 75014 Paris, France
     \and
         GEPI/CNRS UMR 8111, Observatoire de Paris, 5 place Jules Janssen,
         92195 Meudon Cedex, France
     \and
         Observatoire de la C\^ote d'Azur, Dpt. Cassiop\'ee, CNRS-UMR 6202, B.P. 4229, 06304, Nice Cedex 04, France
     \and
         Observatoire de Gen\`eve, Universit\'e de Gen\`eve, 51 ch. des Maillettes, 1290 Sauverny, Switzerland
     \and
         Centro de Astrofísica, Universidade do Porto, Rua das Estrelas, 4150-762 Porto, Portugal
             }

   \date{}

   \abstract {The discovery of the short-period giant exoplanet
     population, the so-called hot Jupiter population, and their link
     to brown dwarfs and low-mass stars challenges the conventional
     view of planet formation and evolution.}
   {We took advantage of the multi-fiber facilities \giraffe\/ and
     \uves\/ (VLT) to perform the first large radial velocity survey
     using a multi-fiber spectrograph to detect
     planetary, brown-dwarf candidates and binary stars.}
   {We observed 816 stars during 5 consecutive half-nights. These
     stars were selected within one of the exoplanet fields of the
     space mission CoRoT.}
   {We computed the radial velocities of these stars and showed that a
     systematic error floor of 30 \ms\/ was reached over 5
     consecutive nights with the \giraffe\/ instrument. Over the whole
     sample the Doppler measurements allowed us to identify a sample
     of 50 binaries, 9 active or blended binary stars, 5 unsolved
     cases, 14 exoplanets and brown-dwarf candidates.  Further higher
     precision Doppler measurements are now necessary to confirm and
     better characterize these candidates.}
   {This study demonstrates the efficiency of a multi-fiber
       approach for large radial-velocity surveys in search for
       exoplanets as well as the follow-up of transiting exoplanet candidates.
       The spectroscopic characterization of the large stellar
       population is an interesting by-product of such missions as
       the \corot\ space mission.}


  \authorrunning{Loeillet B. et al.}
  \titlerunning{Doppler search for exoplanets and binaries using a multi-fiber spectrograph}
  \maketitle

%

\section{Introduction}

The discovery of an ever increasing sample of extra-solar planets (at
present more than 250 objects have been detected) shows
remarkable objects very close to their parent star. This exoplanet population, called the
hot-Jupiter population, is composed of giant planets that revolve at a
short orbital distance around a central star. They are orbiting at
less than 0.1 astronomical units (AU) with an orbital period ranging
from 1.2 to 10 days. Today, more than 40 members of this population
have been detected. This existance of this population suggests
new mechanisms of planet formation and evolution, not envisioned in
the study of our Solar System, such as the migration of the planets in
the proto-planetary disk or gravitational interactions
\citep{Goldreich80,Lin96}.

Mechanisms of planetary formation can be investigated via the mass
function by exploring the gap which separates the high-mass
``planetary'' companions (considered here to be more than M$\sin i =$ 3 \MJ) from
their low-mass ``stellar'' counterparts. The ``super-planet'' or brown
dwarf HD~162020b \citep{Udry02}, the super-massive planet HD147506b
\citep{Bakos07} and the very low-mass stars OGLE-TR-122,123
\citep{Pont05,Pont06} are at the tails of these 2 populations and
illustrate this gap. These suggest two different processes of formation
and evolution of planetary systems. High-mass exoplanets and low-mass stellar companions with short periods
are quite rare. We can only list 6 examples of high-mass
  exoplanets with an orbital period of less than 20 days: Gl86b
  \citep{Queloz00}, Tau Boo b \citep{Butler97}, HIP14810b
  \citep{Wright06}, HD 195019b \citep{Fischer99}, HD 162020b
  \citep{Udry02} and HD147506b \citep{Bakos07}. All of them have a
  minimum mass between 3.5 and 14
  M$_{Jup}$.\\

Ground-based photometric transit surveys such as OGLE
  \citep{Udalski02}, SuperWASP \citep{Pollacco06}, TrES
  \citep{Alonso07}, HAT \citep{Bakos06}, have sufficient
  photometric precision to detect transiting hot Jupiters and
  eclipsing low-mass stars. New space missions, designed to search
for transiting exoplanets, will probe the planetary population at
smaller radii than ever before thanks to an improved photometric
precision, a longer temporal coverage with a better duty cycle, 
and without the observational perturbations induced by the Earth's
atmosphere. In parallel, the next step for radial-velocity (RV)
programs consists of simultaneously observing a large sample of stars
with a reduced on-telescope time. This could be foreseen by using
multi-fiber technology combined with accurate spectrographs.\\

In a first attempt, \citet{Bouchy04,Bouchy05} have demonstrated the
ability of the \uves\/ spectrograph to obtain precise Doppler
measurements. However, only 7 stars could be monitored at the same
time with their instrumental configuration. This may be convenient for
the Doppler follow-up of planetary candidates but is not sufficient
for a large Doppler survey to detect new extra-solar planets
with limited telescope time. With more than 100 fibers distributed
over a large field of view of 25 arcmin, the \giraffe\/ spectrograph
at the VLT \citep{Pasquini00}, used in the {\sl MEDUSA} mode may
overcome this limitation. However, its ability to accurately measure
Doppler shifts has yet to be demonstrated.

In January 2005, we performed a dedicated radial-velocity survey of
more than 800 stars with the \giraffe\/ and the \uves\/
spectrographs. We chose to observe one of the exoplanet fields of the
space mission \corot\/, launched in December 2006 \citep{Baglin03}.
We took advantage of the target characterization of the exoplanet
input catalog which provides an approximate spectral classification
and astrometry of the stars (Moutou et al., 2007, in prep.). Our
scientific goals were to :
\begin{itemize}
\item{ Identify massive hot-Jupiter exoplanet and brown dwarf candidates,}
\item{ Identify binary stars in a \corot\ field, }
\item{ Explore the ability of the multi-fiber instrument \giraffe\/ to accurately measure Doppler shifts
    for more than one hundred stars simultaneously.}
\item{ Check and further improve the spectral classification in
the \corot\/ exoplanet fields by photometric observations.}
\end{itemize}

In this paper, we present the results of our radial velocity analysis
carried out on 816 stars. The spectral analysis of the whole sample of
spectra will be the subject of a separate paper. The sample of stars
is presented in section 2. The data reduction method and the
subsequent RV measurements are detailed in section 3.  The RV
performance is discussed in section 4 and the first results of the
radial velocity analysis are presented in section 5.

\section{Observations}
\subsection{\flames\/ facilities}

In January 2005 we obtained 5 half-nights in visitor mode with the
\flames\ facilities (program 074.C-0633A) attached to the 8.2m Kueyen
telescope (UT2) based at the ESO-VLT. \flames\ is a multi-fiber link
which makes it possible to feed up to 130 targets into the GIRAFFE
echelle spectrograph which covers a 25 arcmin diameter
field-of-view (0.136 \degr$^2$). In addition, a simultaneous
Thorium-Argon calibration is available by using the {\sl MEDUSA} mode
which dedicates 5 fibers to monitor a Th-Ar lamp. The fiber link
allows a stable illumination at the entrance of the spectrograph and
the simultaneous calibration is used to track instrumental drift.

We chose to observe in the HR9B spectral domain, centered at 525.8nm.
This setup covers 200 \AA\/ and a CCD pixel corresponds to 0.05
\AA\/. This instrumental configuration offers the highest resolution
(R=25'800) and the best radial velocity accuracy
\citep[see][]{Royer02}. It indeed provides the best compromise in term
of signal-to-noise and the number of spectral lines necessary for
accurate Doppler measurements.

We combined the \giraffe\ spectrograph with \uves\ which allow us to
simultaneously observe 7 additional stars at a much higher resolution
plus one fiber dedicated to the simultaneous Thorium
calibration. Observations were made with the red arm of the
spectrograph at a central wavelength of 580 nm. This setup covers 2000
\AA\/ with a resolution of 47~000 and a CCD pixel corresponds to 0.015
\AA\/.

\subsection{Target selection and observational strategy}

We chose to observe stars selected in one of the exoplanet \corot\
fields located in a region close to the direction of the Galactic
anti-center.  This field has a homogeneous star density of about 3600
stars per square degree with magnitudes brighter than $r <$ 15. It
will be one of the first fields to be observed by the space mission
during the first year of operation.

We chose 6 different \flames\ fields, with no overlap between each
field, and distributed over an area that covers about 40\%\/ of one
of the two \corot\ exoplanet CCDs. This results in a total of 774
stars observed with the \giraffe\ spectrograph plus 42 additional ones
observed at a higher resolution with \uves.

When selecting our targets in these \flames\ fields, we took advantage
of the \corot\ exoplanet entry catalogue {\it EXODAT}
\citep{Deleuil06}. Built from dedicated ground-based observations and
existing catalogues to prepare the exoplanet program, this catalogue
not only provides the astrometry for all stars within the potential
exoplanet fields of the mission, but also gives an estimate of the
spectral types and luminosity classes of all the stars in the range 11
to 16 in $r$-mag. Thanks to this information, the target selection was
conducted according to 3 different levels of priority, depending on
our scientific goals and on the instrumental constraints: angular
separation between stars for the positioning of the fibers, field of
view of the instrument, etc. Our 3 levels of priorities were defined
as: 1) bright solar-type stars (F, G and K spectral type stars)
and a $V$-mag less than 15; 2) F0 to K5 giant stars and stars with
a $V$-mag greater than 15; 3) O, B and A-type stars, and giant
stars with spectral type ranging from K5 to M.  The \uves\ fibers were
allocated to targets of the first priority only.

  \begin{figure}
\begin{center}
\subfloat[]{
    \includegraphics[scale=0.3]{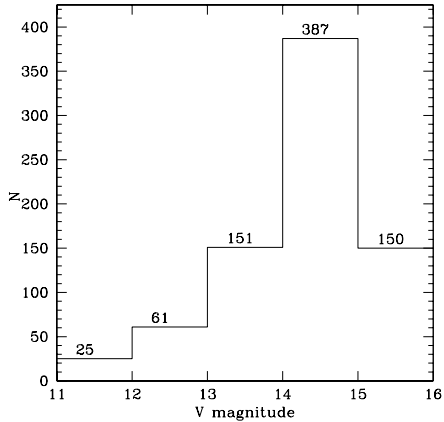}}
\subfloat[]{
     \includegraphics[scale=0.3]{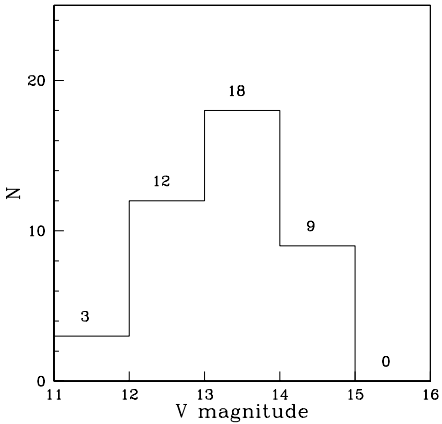}
  }
\caption{Magnitude distribution of the observed stars with the
\giraffe\ (a) and the \uves\ (b) spectrographs.}
\label{distrimag}
\end{center}
\end{figure}

\begin{figure}
\begin{center}
  \subfloat[]{
  \includegraphics[scale=0.4]{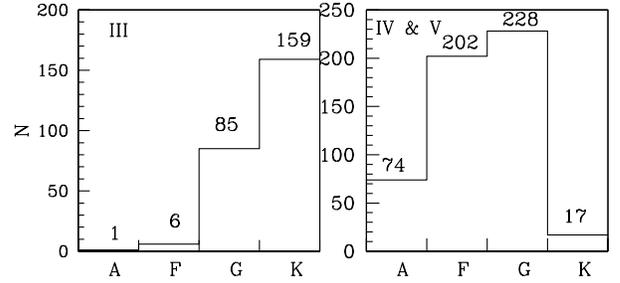}
  }
  \hspace{1cm}
  \subfloat[]{
  \centering{
  \includegraphics[scale=0.4]{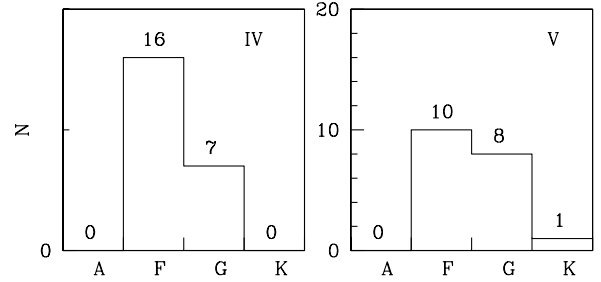}
  }}
\caption{Distribution of the sample of observed stars with the
\giraffe\ (a) and the \uves\ (b) spectrograph over the
spectral type for dwarfs and giants. The sample is completed by a O-type dwarf star
and a M-type giant star.} \label{distriCS}
\end{center}
\end{figure}

Figure~\ref{distrimag} shows the number of stars as a function of $V$
magnitude for the two spectrographs.  Figure~\ref{distriCS} shows the
number of stars as a function of spectral type for the two main
luminosity classes, again for the two spectrographs. The majority of
the fibers have thus been allocated to F, G, K-type dwarfs, which are
most favorable for precise radial velocity measurements. They
represent about 60\% of our targets. The remaining targets are mainly
giants (30\%) and A-type dwarfs (10 \%).

The observational strategy consists of one measurement per night of
each of the 6 fields during the 5 consecutive half-nights with a
typical exposure time of 30 min. One of our fields was observed only 3
times, due to a lack of observational time during 2 of the 5 nights. The journal of the
observations is presented in Table~\ref{journalobs}. We have obtained
5 different spectra for 680 stars and 3 spectra for 136 stars, which
results in a total of 3808 spectra.

\begin{table*}
\caption{\label{journalobs} Journal of the observations with the coordinates
  of the field centers and the exposure time used.}
\begin{center}{
\begin{tabular}{c c c c c c c}\hline
\hline
 & Field 0 & Field 1 & Field 2 & Field 3 & Field 4 & Field 5\\
\hline
$\alpha$ center & 6:42:26 & 6:42:26 & 6:42:26 & 6:44:06 & 6:44:06 & 6:44:06\\\vspace{0.2cm}
$\delta$ center & -1:23:51 & -0:52:30 & -0:27:30 & -1:17:30 & -0:52:30 & -0:27:30\\
MJD - 53393 / & 1.0225 & 1.1865 & 1.0526 & 1.0876 & 1.1203 &
1.1538\\\vspace{0.2cm}
Exposure Time (min) & 30 & 30 & 35 & 35 & 35 & 35\\
 & 2.0228 & - & 2.0651 & 2.1368 & 2.1741 & 2.2041\\\vspace{0.2cm}
 & 49 & - & 30 & 42.3 & 30 & 30\\
 & 3.0249 & 3.1879 & 3.0583 & 3.0912 & 3.1234 & 3.1559\\\vspace{0.2cm}
 & 35 & 25 & 35 & 35 & 35 & 35\\
 & 4.0214 & - & 4.0544 & 4.0877 & 4.1197 & 4.1524\\\vspace{0.2cm}
 & 35 & - & 35 & 35 & 35 & 35\\
 & 5.0194 & 5.1666 & 5.0481 & 5.0774 & 5.1062 & 5.1374\\
 & 30 & 25 & 30 & 30 & 30 & 30\\\hline
\end{tabular}}
\end{center}
\end{table*}

We took care to keep the same configuration
each night. Each star was allocated to the same fiber on the
  instrument on each of the 5 consecutive nights in order to minimize
  errors, which could be due to different chromatic
  behavior of the fibers and/or different pixel responses on the CCD.
  
\section{Data reduction}

In the next sections we describe the essentials of the extraction and
reduction process adopted for the \giraffe\ data. The one used for the
\uves\ data is detailed in \citet{Bouchy04}.

\subsection{Extraction and wavelength calibration}
The frames have been reduced using the \giraffe\ BaseLine Data
Reduction Software (girbldrs v1.12,
see~\citealt{Blecha00,Royer02}). The data were optimally extracted
  to 1D spectra following the original Horne's method
  \citep{Horne86}. Analyzing the Th-Ar calibration spectra, we found
an instability of the zero-point of the absolute wavelength
calibration along the calibrations made during the 5 nights. To
  set the zero-point, we chose to use the Th-Ar calibration spectra
  registered during the first day of our observational campaign. The
  wavelength solution was thus computed once, using this first Th-Ar
  exposure and applied to all the scientific frames. This allowed for
  a better estimation of the instrumental drift between observation
  nights and observational exposures thanks to the simultaneous Th-Ar
  spectra (see Sect. \ref{simuldrift} for more details).  We thus used
  the simultaneous Th-Ar spectra for each stellar observation to
  correct the drift from the first Th-Ar calibration exposure.

\subsection{Radial velocity measurements}
Radial velocities were obtained by a weighted cross-correlation
\citep{Baranne96} of each spectrum with a numerical mask, constructed
from the Solar spectrum atlas \citep{Baranne96,Pepe02}. The resulting
Cross-Correlation Function (CCF) exhibits a Gaussian shape. By fitting
it by a Gaussian function, we derived the full width half maximum
(FWHM in \kms) which gave an estimate of the projected rotational
velocity of the star, the depth (C in \%\/) and the radial velocity
(RV in \kms) given by the center of the Gaussian. In our study, we
used the same G2 template mask for the \uves\ and the \giraffe\
spectra. It is adapted for most solar-type stars. We performed the
same analysis with different masks which gives no significant
improvements on the RV determination (see
Sect.~\ref{RVanalysis}). Finally we corrected the calculated RV by the
barycentric Earth RV.

\subsection{\label{photonoise} Photon noise analysis}
Our RV measurements are not photon-noise limited. They depend on the photon noise and an instrumental error floor set by systematic effects. This error floor clearly appears for stars with a short RV variation, as illustrated by Fig.~\ref{dispbphot}. Small observed RV dispersion (RMS) values are indeed set around an average value much larger than 0 \ms. The floor error is set by the asymptotic limit for these stars where the photon noise has very little impact. We estimated this error floor to be equal to 30 {\ms\/}. The same level of systematic effects was found for \uves\ \citep[see][]{Bouchy05}.
Following the analysis carried out by \citet{Bouchy05}, we
computed the photon noise uncertainties through the following
relation:
$$\sigma_{RV} = \sigma_0 \ast \frac{\sqrt{FWHM}}{SNR \ast C}$$
where $SNR$, $FWHM$ and $C$ are respectively the signal-to-noise ratio
per pixel, the full width half maximum and the depth of the CCF
peak. The constant $\sigma_0$ was determined empirically. To evaluate the impact of the photon noise and the value of
$\sigma_0$ for  the \giraffe\/ spectrograph, one must move into the regime of large RV RMS, where the floor error only enters weakly.
The $\sigma_0$ values were adjusted so that for the high RV
  dispersion values the photon noise uncertainty statistically matches
  the measured RV RMS dispersion. It is equal to 3 for \uves\ and to 10.5 for \giraffe. These values are related to
the difference in resolution, spectral coverage and the total efficiency of the spectrographs. 

\subsection{\label{simuldrift} Simultaneous instrumental drift correction}
For the whole set of spectra, we computed the simultaneous
instrumental drift of a single exposure, using the 5 Th-Ar fibers.  We
found that the two extreme Th-Ar fibers, located near the edges of the
\giraffe\ CCD, present a significant drift departure, indicating that
the wavelength calibration provides less constraint on fibers at the
edge on the CCD. We thus used the averaged value of the 3 central
Th-Ar fibers of each individual exposure to correct the instrumental
drift of the individual exposure with the Th-Ar wavelength
calibration. Typically the correction applied is about few
  hundred {\ms}. We checked that the correction made with the average
value of the 5 Th-Ar fibers does not drastically change the drift
correction in our whole analysis.

\subsection{\label {bglight} Background light correction}
Spectra acquired during the first three nights were contaminated by
the background light due to the Moon. In particular, the Moon produces
an additional peak in the CCF which affects RV measurement when the
star's RV is close to the Moon's. We found that about 20\%\ of our
targets are affected in this way. We defined the
following procedure to correct for the Moon light in our spectra. For
each \flames\ field we selected one spectrum that presents a clear
background Moon light and a low SNR stellar spectrum with very shallow
spectral lines. We then subtracted this spectrum from all the spectra
obtained during the same exposure. We checked that this correction
clearly allows us to reduce the RV dispersion for the sample of
affected targets. However the correction is not efficient for small
RV variations (less than 200~\ms). Indeed we noted that the difference
between the calculated barycentric correction of the two extreme
fibers positioned from the East to the West and distant by
about 25 arcmin, reaches 200~\ms\ for the same exposure. The observed background 
light component is then affected in a similar way. The corresponding peak in the CCF is thus shifted in RV.
This demonstrates that a single fiber dedicated to sky monitoring is not
sufficient. However, we will see in Sect.~\ref{RVanalysis} that
this only affects a small fraction of the RV variable stars. One
solution to optimize the sky monitoring would be to allocate
several fibers in the $\alpha$ direction to limit the difference of
the calculated barycentric correction between sky-monitoring fibers
and scientific fibers.

\section{\label{perf} Radial velocity performance}

\begin{figure}
\begin{center}
\includegraphics[scale=0.45]{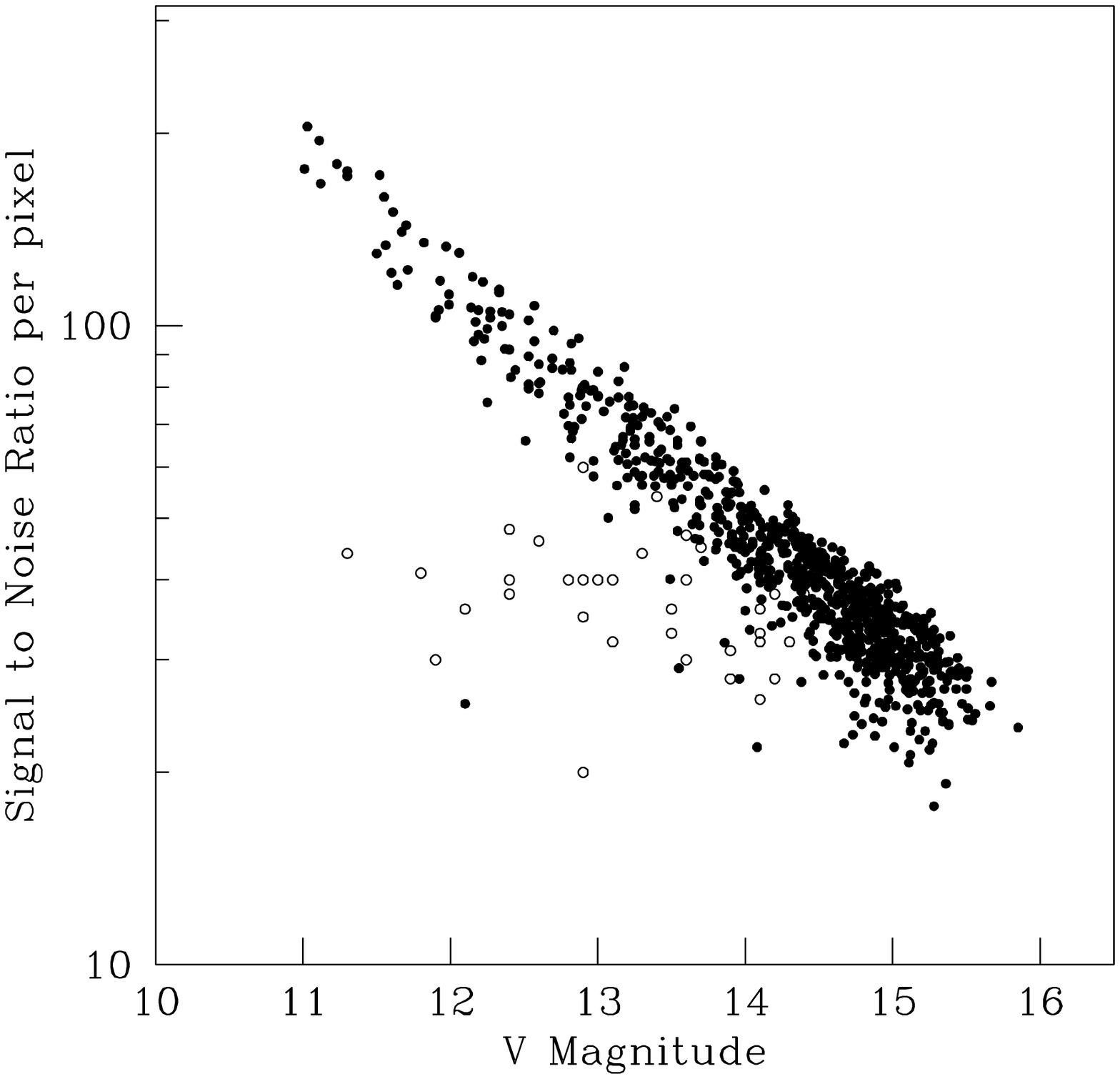}
\caption{\label{sn}Signal-to-noise ratio per pixel of all the spectra
  measured on the last observation night as a function of stellar
  magnitude. The \giraffe\ data is shown by filled circles (1 pixel
  $=$ 0.05 \AA\/) and the \uves\ data is shown by open circles (1
  pixel $=$ 0.015 \AA\/). The signal-to-noise scale is a logarithmic
  scale and we note that both parameters are linked by a clear
  logarithmic relation.}
\end{center}
\end{figure}

The CCF of 115 stars, representing about 14\% of the whole sample,
exhibits no significant correlation peak. Different kind of situations
could generate a no-CCF: i$)$ spectra with too small a SNR, ii$)$ stars
with a very large $v\sin{i}$ (typically $>$ 50 \kms), iii$)$ early
type stars whose stellar features do not match the template well and
iv) multiple-line spectroscopic binaries whose resulting CCF
peaks of each component are blended, reducing the contrast of their
peak. We checked that among this set of targets with no CCF peak,
there is no correlation with the magnitude of the target nor its
position within the slit of the instrument. The spectral type of
almost all these stars is in the range F8 to O9 and thus they are
likely to be rapidly rotating \citep{Allen00}. We computed the CCF for
these specific stars with an F0 template mask which gave no
improvement. Our dynamical radial velocity study was thus restricted
to the stars yielding a CCF peak. We also narrowed the sample to the
stars observed 5 times either with the \giraffe\ or the \uves\
spectrographs. This results in a total of 584 targets
suitable for our radial velocity analysis
(Table~\ref{totalbudget}).\\

\begin{table}[h]
\caption{\label {totalbudget} Sample of stars observed 3
  or 5 times.}
\begin{center}{
\begin{tabular}{c c c c c}\hline
\hline
 & \multicolumn{2}{c|}{\giraffe\ targets} & \multicolumn{2}{c}{\uves\ targets}\\
 & With & No & With & No\\
 & CCF & CCF & CCF & CCF \\
\hline
5 meas. & 550 & 95 & 34 & 1\\
3 meas. & 110 & 19 & 7 & 0\\
Total & 660 & 114 & 41 & 1\\
\hline
\end{tabular}}
\end{center}
\end{table}

\begin{figure*}
  \centering \subfloat[Distribution of the RV dispersions of the observed stars as a function
  of their average photon-noise uncertainty. Both scales are logarithmic. The filled
  circles represent the dispersions measured with the \giraffe\
  instrument and the open circles represent the ones obtained with the
  \uves\ instrument. The $\sigma$ curve represents the estimated detection limit at 1$\sigma$ of the studied sample of stars.
  The $2.1*\sigma$ curve shows our threshold of detection to detect RV variable stars
  It is derived from simulations presented
  in Fig.~\ref{simustatdisp}. Stars with identified composite spectra
  (Multiple spectroscopic binaries, see~\ref{SB}) have been removed
  from this figure.]{
    \label{dispbphot}
    \includegraphics[width=8cm]{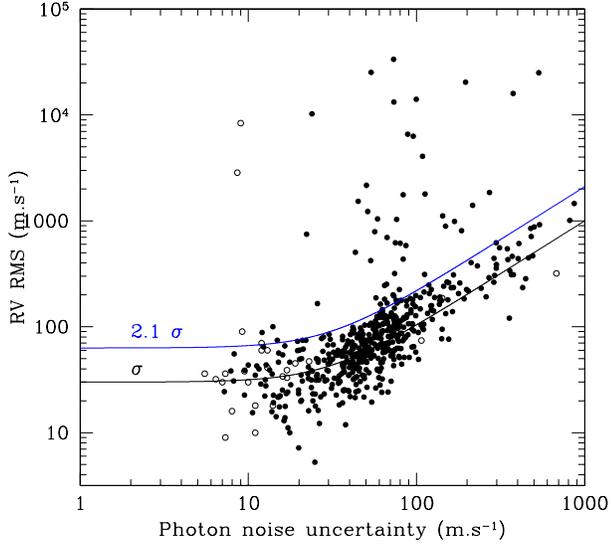}
  } \hspace{1cm} \subfloat[Distribution of simulated dispersions of 5 RV
  as a function of the calculated photon noise uncertainty on the observed stars.
  As in Fig.~\ref{dispbphot} the $\sigma$ curve represents the estimated detection limit at 1$\sigma$ of the studied sample of stars.
  The $2.1*\sigma$ shows our detection threshold.]{
    \includegraphics[width=8cm]{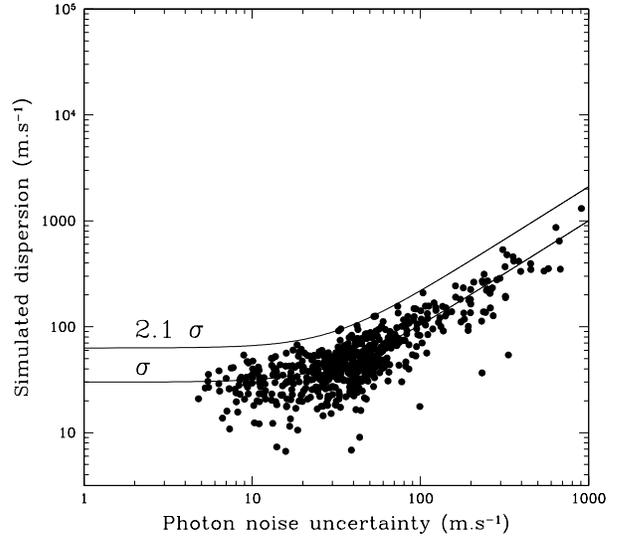}
    \label{simustatdisp}
  }
  \caption[width=17cm]{Distribution of the RV dispersion as a function
    of photon-noise uncertainty for the \giraffe\ targets.}
\end{figure*}

We computed the dispersion of the 5 RV measurements and plotted them
as a function of the average photon noise
uncertainty. Figure~\ref{dispbphot} displays the RV dispersion (RMS)
as a function of the estimated photon noise uncertainty for each of
these stars. We have excluded stars with composite spectra (multiple
spectroscopic binaries). The photon noise uncertainties range from 7
to 700 \ms and the RV dispersion ranges from 5 \ms\ to 100 \kms.\\

In order to distinguish the real RV variations of stars from
variations generated by statistical effects, we performed a
Monte-Carlo simulation. We generated 5 simulated RV for each value of
the measured photon noise uncertainty. For that purpose, we used 2
Gaussian distributions with a mean value of zero and a standard deviation
equal to the photon noise for the first one and to a systematic error of 30 \ms\/ for the second one. We then computed the corresponding dispersion of these 5
random values. The results of these simulations are shown in
Fig.~\ref{simustatdisp}. They show that a threshold of 2.1 $\sigma$
corresponds to one false detection due to statistical effects over the
whole sample.

\section{\label{RVanalysis} Radial velocity variation analysis}

Among the set of 701 stars with detected peaks in the CCF and observed
3 or 5 times, we identified 23 spectroscopic binary stars with multiple
spectral components in their CCF. Among the set of 584 stars with CCF
peak and observed 5 times with the \giraffe\ and \uves\ spectrographs,
our dynamical analysis (see Fig.~\ref{dispbphot}) allows us to
identify 66 targets with a RV dispersion greater than the 2.1
$\sigma$ threshold.  These targets include spectroscopic binaries of
type 1 (SB1), blended binary stars (two merged CCF peaks), active stars
and low-mass companion candidates (brown dwarf and exoplanet
candidates). The correction of the background light due to the Moon is
not totally efficient and some stars are still affected as described
in Sect.~\ref{bglight}. We identified and removed 13 of these 66
targets that are affected by the incomplete Moon light
correction. For the remaining 53 stars we looked for an orbital
solution by fitting the 5 RV measurements with a sinusoidal curve.  We
derived the semi-amplitude K, the period P, the systemic velocity
V$_0$, and the phase T$_0$. Five RV measurements are insufficient to
fit an eccentric orbit, therefore we assumed circular orbits. This
assumption is justified for orbital periods shorter than 5 days, as
the orbit is expected to be circularized by tidal effect
\citep{Halbwachs05,Marcy05}. To compute an estimate of the minimum
mass of the companion, we derived the stellar masses of these targets
using the standard tables of stellar masses \citep{Gray88} and the
spectral type from the \corot\ exoplanet catalogue.

\subsection{\label {SB} Spectroscopic binary stars}
The so-called spectroscopic binary (SB2) objects show two distinct peaks in their CCF,
related to the blended spectra of the two components. Twenty-two SB2
and one clear triple system have been identified. Three SB2 do not
present RV variation of both components, probably indicating
long-period binary stars.

We also identified 27 SB1, defined as targets with an estimated
companion minimum mass that is greater than the canonical sub-stellar
limit of 70 \MJ\/ \citep{Chabrier00}. Two of those identified in the
field were observed only 3 times. They present an RV variation greater
than 4 \kms\/. Table~\ref{sbtable} presents the characteristics of the
central star and a rough estimation of the orbital period of the
stellar companion. For periods greater than 20 days, we only give an
estimated period range.

\begin{table*}
\caption{\label{sbtable} Binary systems.}
\begin{center}{
\begin{tabular}{c c c c c c}\hline
\hline
CoRoT & $\alpha$ & $\delta$ & m$_V$ & Binary type & Estimated \\
ID & (h mn s) & (d mn s) &  &  & period range\\
 & (J2000) & (J2000) &  &  & (days)\\\hline
102613769 & 6 41 45.38 & -1 15 36.65 & 15.01 & SB1 & P $>$ 50 d\\
102614070 & 6 41 45.77 & 0 30 53.32 & 14.98 & SB2 & P $\sim$ 20 d\\
102619385 & 6 41 53.58 & -1 23 36.49 & 14.62 & SB2 & P $\sim$ 8 d\\
102619435 & 6 41 53.64 & -0 54 05.29 & 14.69 & SB1 & Unsolved P (3 meas.)\\
102627314 & 6 42 05.32 & -0 29 48.73 & 14.63 & SB1 & P $>$ 100 d\\
102627556 & 6 42 05.7 & 0 23 25.19 & 14.67 & SB2 & P $>$ 50 d\\
102629764 & 6 42 08.98 & 0 31 30.29 & 14.11 & SB1 & P $<$ 20 d\\
102631863 & 6 42 12.06 & -1 28 26.65 & 14.72 & SB1 & 20 d $<$ P $<$ 50 d\\
102634388 & 6 42 16.01 & -0 49 02.82 & 13.55 & SB2 & Unsolved P (3 meas.)\\
102637099 & 6 42 19.43 & -0 53 00.82 & 14.25 & SB2 & Unsolved P (3 meas.)\\
102639561 & 6 42 22.68 & -1 27 01.30 & 11.93 & SB2 & P $>$ 100 d\\
102646279 & 6 42 31.39 & -0 49 59.05 & 12.41 & SB2 & P $>$ 100 d\\
102648472 & 6 42 34.22 & -0 34 21.94 & 14.49 & SB1 & Unsolved P (5 meas.)\\
102648907 & 6 42 34.76 & -1 04 31.22 & 14.53 & SB1 & Unsolved P (3 meas.)\\
102651332 & 6 42 37.86 & 0 24 30.28 & 15.36 & SB1 & 20 d $<$ P $<$ 50 d\\
102651632 & 6 42 38.21 & 0 30 33.66 & 14.12 & SB2 & P $\sim$ 15 d\\
102652553 & 6 42 39.41 & 0 38 56.29 & 14.89 & SB2 & P $\sim$ 12 d\\
102662997 & 6 42 52.63 & -1 10 10.88 & 12.84 & SB1 & P $\sim$ 9 d\\
102663892 & 6 42 53.77 & -1 17 53.56 & 12.4 & SB1 & 20 d $<$ P $<$ 50 d\\
102668497 & 6 42 59.60 & 0 25 00.91 & 13.30 & SB1 & P $\sim$ 11 d\\
102669579 & 6 43 00.93 & 0 26 23.93 & 15.00 & SB1 & P $\sim$ 14 d\\
102672236 & 6 43 04.28 & -1 20 28.28 & 15.07 & SB2 & P $\sim$ 12 d\\
102673596 & 6 43 06.00 & -1 16 08.00 & 13.69 & SB2 & P $>$ 100 d\\
102677302 & 6 43 10.53 & -1 12 00.90 & 13.04 & SB1 & P $\sim$ 11 d\\
102683896 & 6 43 18.80 & -1 16 41.74 & 14.61 & SB2 & P $>$ 50 d\\
102686545 & 6 43 22.12 & -1 12 34.70 & 13.96 & SB1 & P $>$ 50 d\\
102692263 & 6 43 29.23 & 0 56 33.72 & 14.97 & SB1 & P $\sim$ 8 d\\
102694802 & 6 43 32.34 & -1 19 20.35 & 14.47 & SB1 & 20 d $<$ P $<$ 50 d\\
102708308 & 6 43 48.88 & 0 59 08.09 & 15.15 & SB1 & P $\sim$ 18 d\\
102708916 & 6 43 49.76 & 0 47 51.25 & 13.97 & SB2 & P $\sim$ 5 d\\
102709159 & 6 43 50.10 & 0 41 31.02 & 14.00 & SB1 & P $\sim$ 17 d\\
102709968 & 6 43 51.21 & 0 34 54.01 & 14.32 & SB2 & P $\sim$ 12 d\\
102712875 & 6 43 55.22 & 0 51 09.00 & 14.38 & SB2 & P $\sim$ 16 d\\
102713079 & 6 43 55.51 & 0 46 26.69 & 14.73 & SB1 & 20 d $<$ P $<$ 50 d\\
102715243 & 6 43 58.47 & -1 00 51.62 & 13.1 & SB2 & P $>$ 100 d\\
102715978 & 6 43 59.54 & -1 29 34.30 & 13.13 & SB2 & P $\sim$ 4 d\\
102716305 & 6 43 59.97 & 0 44 10.32 & 14.56 & SB1 & P $\sim$ 18 d\\
102718650 & 6 44 03.33 & -1 21 24.66 & 13.7 & SB2 & P $\sim$ 16 d\\
102718810 & 6 44 03.56 & 0 57 42.80 & 13.32 & SB1 & 20 d $<$ P $<$ 50 d\\
102720035 & 6 44 05.35 & -1 15 26.89 & 13.57 & SB1 & P $\sim$ 20 d\\
102725454 & 6 44 13.08 & -1 17 36.53 & 13.35 & SB2 & P $>$ 100 d\\
102726103 & 6 44 14.30 & -1 13 44.44 & 14.54 & SB2 & P $\sim$ 4 d\\
102732890 & 6 44 23.49 & -1 11 31.70 & 13.79 & SB1 & P $>$ 50 d\\
102734591 & 6 44 25.75 & -1 17 29.40 & 14.39 & SB1 & P $\sim$ 8 d\\
102737852 & 6 44 30.25 & 0 41 54.13 & 15.11 & SB1 & P $>$ 50 d\\
102738614 & 6 44 31.29 & -1 10 20.86 & 14.45 & SB2 & P $\sim$ 7 d\\
102740955 & 6 44 34.42 & -1 19 9.12 & 13.27 & SB2 & P $\sim$ 8 d\\
102747222 & 6 44 42.72 & 0 45 22.61 & 12.89 & SB1 & 20 d $<$ P $<$ 50 d\\
102748346 & 6 44 44.27 & -1 20 54.64 & 15.07 & SB1 & P $\sim$ 15 d\\
102748356 & 6 44 44.28 & -1 18 08.24 & 12.9 & SB3 & P $\sim$ 5 d\\
\hline
\end{tabular}}
\end{center}
\end{table*}

A complete analysis of this sample of binary systems will be completed 
thanks to complementary photometric observations and RV
measurements. Indeed the BEST \citep{Rauer04} instrument has
already observed a large part of the \corot\ fields and will probably
bring additional constraints on these systems. In parallel the
space mission \corot\ \citep{Baglin03} will observe this field by the
end of 2007.

\subsection{\label {TsVRv} Targets with small RV variations}

Besides the SB1 stars, a subset of 28 stars exhibit an RV dispersion
above the 2.1 $\sigma$ detection threshold, which corresponds to
63~\ms\ for small photon-noise uncertainty. According to the orbital
solutions that we found, sub-stellar companions may induce these RV variations. We checked the origin of these RV variations by performing
a bisector analysis of the CCF profiles. This allows us to disentangle
RV variations due to true companion from stellar activity.

\subsubsection{Possible blended binary stars and stellar activity signature}

\begin{table*}
\caption{\label{cdtsbissec} Characteristics of identified targets
with a line-bisector effect. The proposed interpretation is based on CCF mask analysis.}
\centering{
\begin{tabular}[width=17cm]{c c c c c}\hline
\hline
ID & $\alpha$ & $\delta$ & m$_V$ & Proposed interpretation\\
COROT & (h mn s) & (d mn s) &  & \\
 & (J2000) & (J2000) &  &  \\
\hline
102619034 & 6 41:53.06 & -0 20 05.35 & 14.95 & Activity\\
102631629 & 6 42 11.71 & 0 29 53.27 & 13.87 & Blended Binary\\
102631928 & 6 42 12.17 & 0 28 38.50 & 15.39 & Activity\\
102700329 & 6 43 39.10 & -0 19 58.87 & 12.81 & Activity\\
102700855 & 6 43 39.69 & 0 31 05.84 & 14.58 & Blended Binary\\
102717173 & 6 44 01.22 & 0 27 56.92 & 14.34 & Blended Binary\\
102723949 & 6 44 10.95 & -1 11 13.24 & 13.96 & Activity\\
102724171 & 6 44 11.26 & -1 10 09.88 & 13.43 & Blended Binary\\
102743523 & 6 44 37.80 & -0 34 17.98 & 11.50 & Blended Binary\\
\hline
\end{tabular}}
\end{table*}

By analyzing the CCF profiles of these targets with a bisector
analysis we found that 9 of them presented line-bisector
variations. These variations could either be induced by a background
blended binary \citep{Queloz01} or by stellar photometric activity
\citep{Santos02}. We completed our analysis by exploring the behavior of the RV RMS of these stars when cross-correlated their spectra using 4 different masks. We used masks constructed from M4, K0, K5 and F0 type stars.
If the RV RMS variation is significant we consider the target to be a
blended binary system. On the other hand if no variation is found,
the stellar activity origin of the RV variation is more appropriate. We thus identified 5 possible blended binary star
systems which present a cross-correlation mask effect and 4 active
stars (Table~\ref{cdtsbissec}).

\subsubsection{\label{candidates} Substellar companion candidates}
\begin{figure*}
\centering
\subfloat[]{
\includegraphics[scale=0.7]{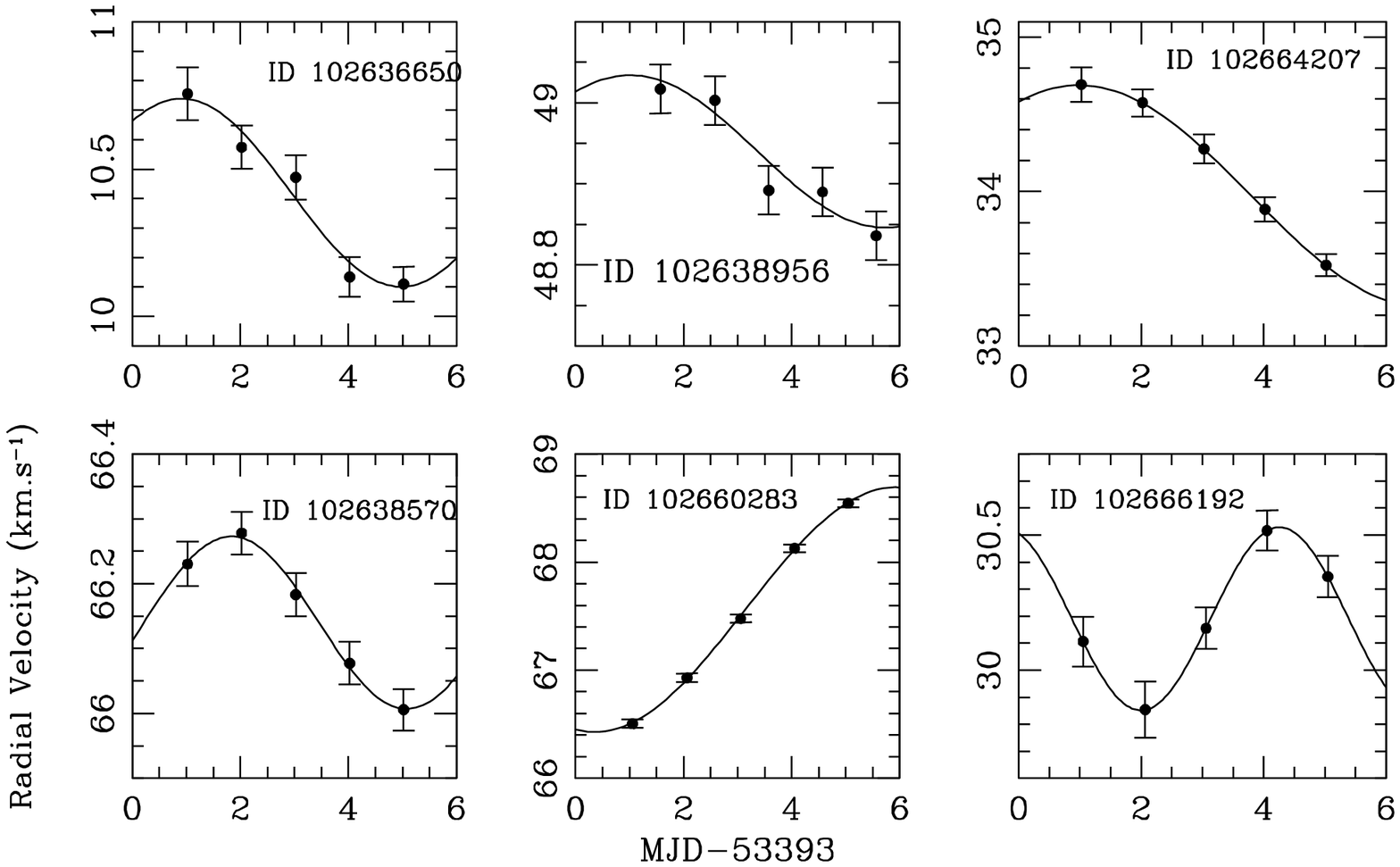}}
\subfloat[]{
\includegraphics[scale=0.7]{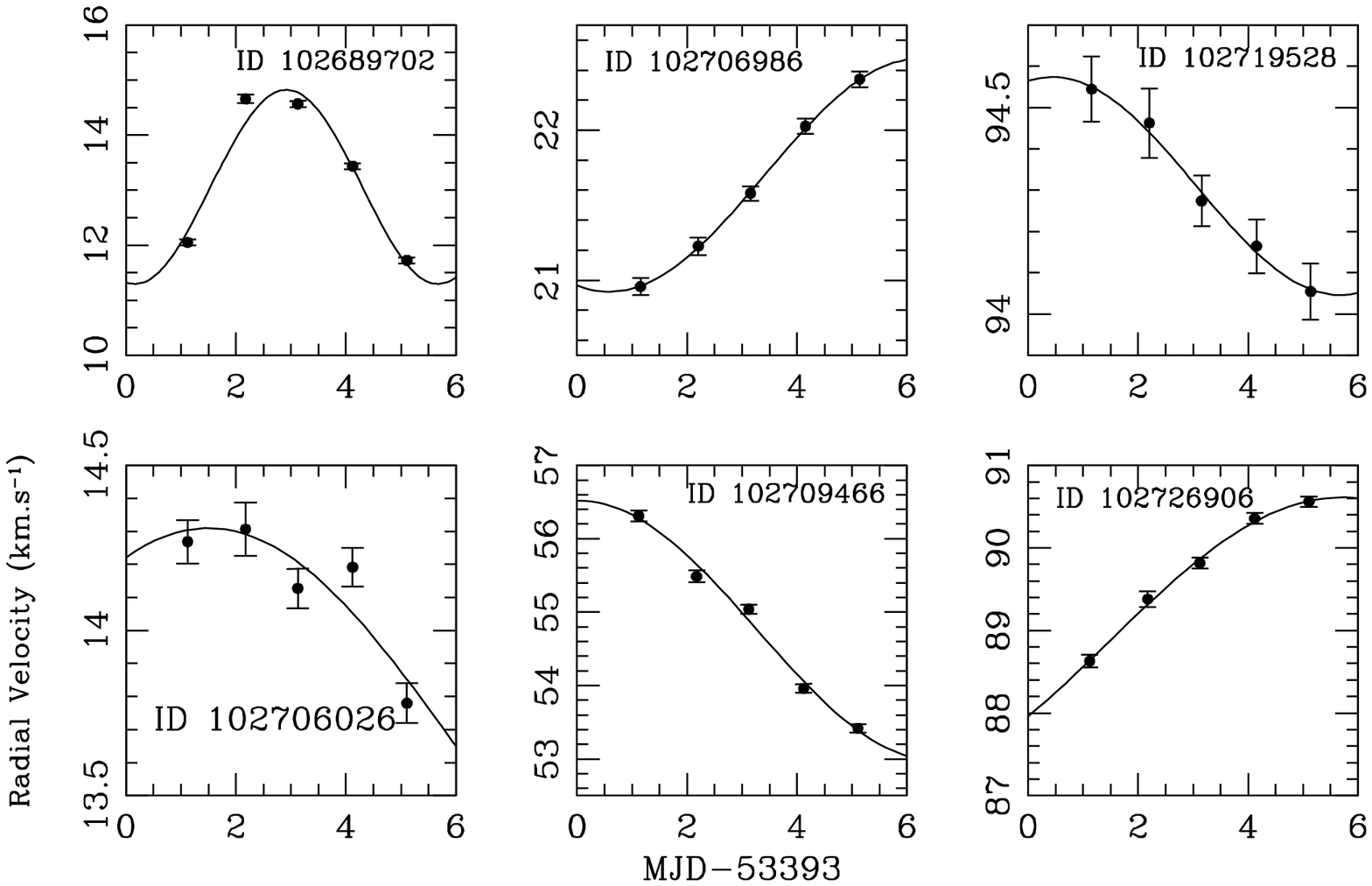}}
\subfloat[]{
\includegraphics[scale=0.5]{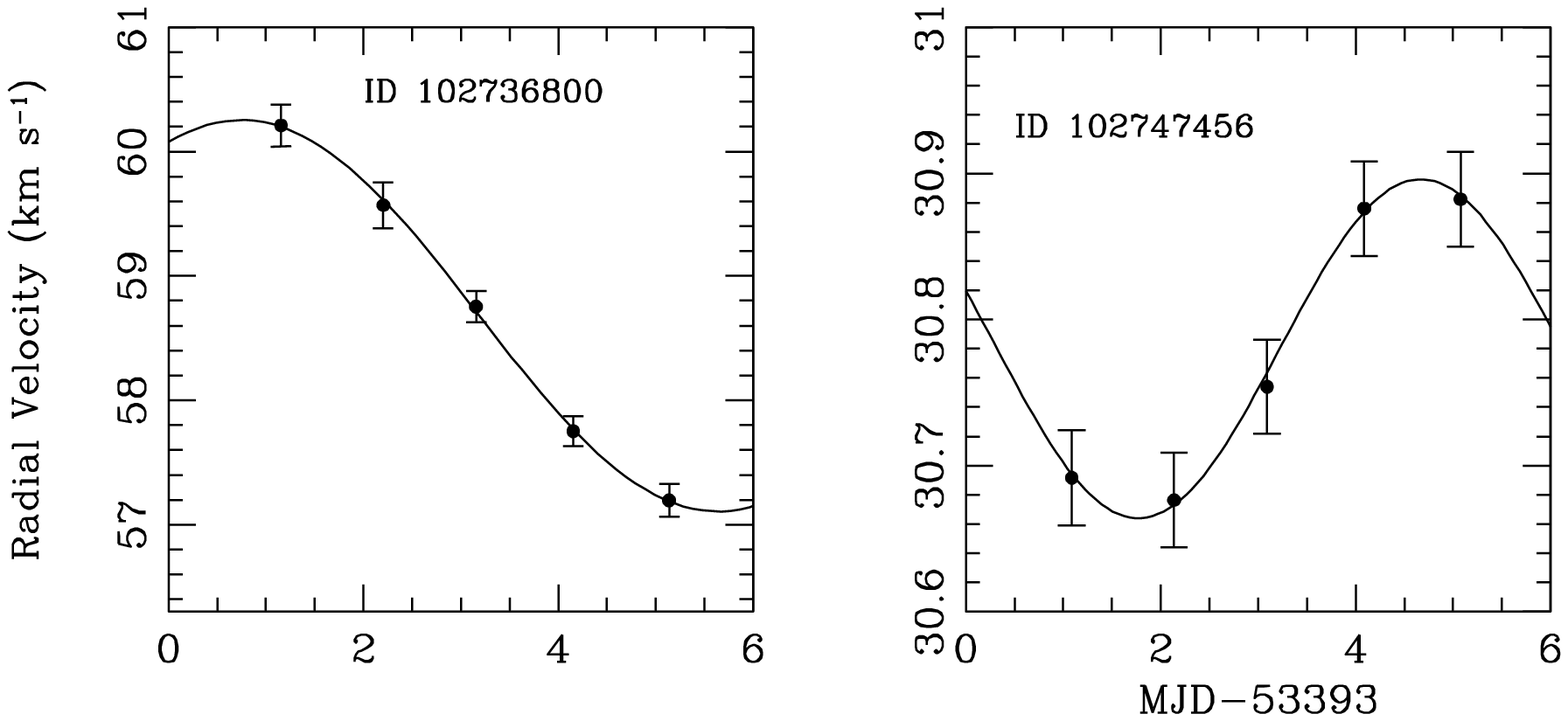}
}
\caption{\label {fig_cdts} Sinusoidal fits of the 14 exoplanet and
  brown dwarf candidates. The RV measurements are represented with
  their error bars and the curve represents the best sinusoidal fit.}
\end{figure*}

Fourteen candidates present an orbital solution that matches the
exoplanet and brown dwarf mass category and that do not show a
line-bisector effect nor cross-correlation mask effects.
Figure~\ref{fig_cdts} shows the best orbital solution we obtained for
those 14 candidates. Tables~\ref{scaractscdts} and \ref{orbparacdts}
present the parameters of the central stars and the estimated orbital
parameters.

A detailed spectral analysis of the stellar spectra of the whole
sample is currently under-progress and will be presented in a
forthcoming paper. We used on our
substellar companion candidate spectra an automatic spectral analysis
algorithm \citep{RecioBlanco06}, developed for the GAIA/RVS
spectroscopic instrument \citep{Wilkinson05}. This also constitutes the preparation for a planned
follow-up program aiming at deriving fundamental parameters of the
dwarf population in the \corot\ exoplanet fields. This algorithm, MATISSE(MATrix Inversion for Spectral
SynthEsis), derives the stellar parameters and measures the abundances
of elements present in the spectra.
It employs a grid of synthetic spectra covering the observed spectral
domain and computed with the same atmosphere models and line lists
\citep[see details in][]{RecioBlanco06}. For our RV study, we checked
that the spectral classification (Table~\ref{cdtsbissec} and
\ref{scaractscdts}) of our subset of candidates derived from the
photometric observations on the one hand and from spectral analysis on
the other hand are in agreement. We used the spectral type
determination made with MATISSE to calculate the minimum mass of our
sub-stellar companion candidates.

\begin{table*}
  \caption{\label{scaractscdts} Brown dwarf and exoplanet
    candidate hosting stars. The spectral types and luminosity classes
    are determined from our spectral analysis carried out with MATISSE and with the photometric analysis.}
\centering{
\begin{tabular}[width=17cm]{c c c c c c}\hline
\hline
\corot\ & $\alpha$ & $\delta$ & m$_V$ & Spectral type & Spectral type\\
ID & (h mn s) & (d mn s) &  & of the parent star & of the parent star\\
 & (J200) & (J2000) &  & (photometric analysis) & (MATISSE analysis)\\
\hline
102636650 & 6 42 18.87 & -1 24 06.12 & 15.16 & K0V & G8V\\
102638570 & 6 42 21.35 & -1 11 47.58 & 12.27 & K4III & K3III\\
102638956 & 6 42 21.88 & -0 31 50.74 & 13.52 & G2V & G2IV\\
102660283 & 6 42 49.20 & 0 17 59.78 & 12.17 & K3III & G8III\\
102664207 & 6 42 54.18 & -1 20 37.79 & 15.15 & F6V & F8V\\
102666192 & 6 42 56.67 & 0 19 22.08 & 14.62 & G2V & F5V\\
102689702 & 6 43 25.98 & 0 48 32.9 & 13.90 & F5V & F8V\\
102706026 & 6 43 45.86 & -0 50 31.60 & 14.45 & G2V & F8V\\
102706986 & 6 43 47.08 & 0 23 50.14 & 14.04 & G2V & F8V\\
102709466 & 6 43 50.51 & 0 50 03.52 & 14.80 & G2V & G1V\\
102719528 & 6 44 04.62 & 0 36 31.57 & 15.19 & G0III & G0IV\\
102726906 & 6 44 15.39 & 0 53 43.22 & 15.39 & G0III & G7IV\\
102736800 & 6 44 28.79 & 0 35 21.73 & 15.25 & G0III & No output\\
102747456 & 6 44 43.05 & -1 22 13.84 & 11.90 & K4III & K3III\\
\hline
\end{tabular}}
\end{table*}

\begin{table*}
\caption{\label{orbparacdts} Estimated orbital parameters of the brown dwarf and exoplanet
  candidates, average photon-noise uncertainty and residuals of the sinusoidal
  and linear fits. We quadratically
  added systematic errors of 30 {\ms} to obtain the global error on
  the measurements. The estimated mass of the parent star based on the MATISSE
  analysis and the companion candidate are also given. The targets in bold
  type are our best candidates obtained by comparing the residuals from both solutions.}
\centering{
\begin{tabular}[width=17cm]{c c c c c c c c c}\hline
\hline
\corot\ & K & P & V$_0$ & Average & Residuals from & Residuals from & Estimated mass & Estimated \\
ID & (\kms) & (days) & (\kms) & photon-noise uncertainty & a sinusoidal fit & a linear fit & of the parent star & M$_2$ $\sin(i)$\\
 &  &  &  & (\ms) & (\ms) & (\ms) & (M$_{\odot}$) & (\MJ)\\\hline
102636650 & 0.32 & 8.1 & 10.42 & 44.3 & 47.1 & 57.7 & 0.9 & 3.0\\
102638570 & 0.13 & 6.5 & 66.14 & 9.3 & 6.6 & 39.3 & 2.2 & 2.1\\
102638956 & 0.09 & 9.4 & 48.9 & 14.1 & 19.5 & 21.0 & 1.0 & 1.0\\
{\bf 102660283} & 1.13 & 11.2 & 67.56 & 14.8 & 16.7 & 55.8 & 2.4 & 23.4\\
102664207 & 0.72 & 11.2 & 33.97 & 55.5 & 0.7 & 75.8 & 1.2 & 9.4\\
{\bf 102666192} & 0.34 & 4.5 & 30.19 & 52.8 & 0.8 & 156.3 & 1.3 & 3.4\\
{\bf 102689702} & 1.76 & 5.5 & 13.01 & 34.2 & 231.8 & 1200 & 1.2 & 18.1\\
102706026 & 0.6 & 18.0 & 13.7 & 38.2 & 80.0 & 108.2 & 1.2 & 8.7 \\
102706986 & 0.78 & 11.2 & 21.70 & 28.9 & 8.9 & 44.0 & 1.2 & 10.2\\
102709466 & 1.77 & 13.1 & 54.76 & 39.1 & 110.4 & 113.9 & 1.1 & 22.9\\
102719528 & 0.26 & 10.5 & 94.31 & 43.1 & 15.4 & 22.8 & 1.6 & 4.0\\
{\bf 102726906} & 1.74 & 16.9 & 89.04 & 44.5 & 40.8 & 120.3 & 1.5 & 30.2\\
102736800 & 1.57 & 9.8 & 58.68 & 95.4 & 27.5 & 101.2 & 2.1 & 28.4\\
102747456 & 0.12 & 5.8 & 30.78 & 8.0 & 4.6 & 32.6 & 2.2 & 1.8\\
\hline
\end{tabular}}
\end{table*}

We checked the likelihood of our 14 candidates by comparing the
residuals of the sinusoidal and linear fits to the global uncertainty
(photon-noise uncertainty quadratically added to the systematic
error). A linear fit would suggest a long-period orbiting object.  In
most cases, the residuals of the sinusoidal and linear fits are
comparable to, or even less than the global uncertainty. Three
candidates exhibit a residual much larger than the global uncertainty
which could be the signature of an eccentric orbit or stellar
activity. The results are presented in Table~\ref{orbparacdts}. It is
statistically difficult in most cases to disentangle the sinusoidal
and linear solutions. They are both compatible with their global
uncertainty. However 4 candidates, namely \corot\/ ID 102660283,
102666192, 102726906 and 102689702 (in bold type in
Table~\ref{orbparacdts}), present significant sinusoidal solutions that are
statistically better than a linear solution. They appear as our best
candidates. One is in the planetary mass range and 3 are in the BD
mass range. Taking into account the BD desert \citep{Grether06}, our
BD candidates may also be longer-period stellar companions.

\subsubsection{Unsolved orbital solution}
From the sample of 28 stars, 5 targets remain above the detection
threshold with small RV variations and unsolved orbital solutions.

Two of them, \corot\ ID 105936100 and 102724641, do not present any
satisfactory orbital solutions and their RV variations could be
related to stellar activity.  The three other targets, namely \corot\/
ID102622204, ID102638630 and ID102653533, exhibit large variations of
the contrast of their CCF. We checked that these variations are not
due to a variation of the SNR nor to fiber-to-fiber
contamination. Using a cross-correlation mask analysis, we ruled out
a blended binary scenario as the origin of this phenomena.

\section{Discussion}

\subsection{Detection capability of the GIRAFFE instrument}
We performed Monte-Carlo simulations in order to illustrate the
capability of the \giraffe\ spectrograph to detect binary stellar systems and
exoplanets with 5 consecutive RV measurements. For different
stellar companion masses and periods and a solar-mass parent star,
we computed the RV of 5 consecutive dates, separated by one day,
  for which the first one was randomly initiated over the period with
  a random initial phase. We quadratically added a random noise to the
  systematic error. This noise corresponds to the estimated photon
  noise uncertainty given a detection threshold thanks to the $\sigma$
  curve determined previously for the \giraffe\/ instrument. This
  photon noise uncertainty corresponds to the value on the x-axis of
  the $\sigma$ curve for the defined detection threshold on the
  y-axis. We then computed the dispersion and estimated the
  detection probabilities of such systems as a function of the
  detection threshold. Only circular orbital cases were
  considered and we did not include a distribution of orbital
  inclinations. In these simulations, we only considered the induced
  RV motion of a star that a planet implies and compared it to the
  detection threshold obtained for the \giraffe\/ instrument.

\subsubsection{For binary systems}
\begin{figure}
\begin{center}
\includegraphics[scale=0.45]{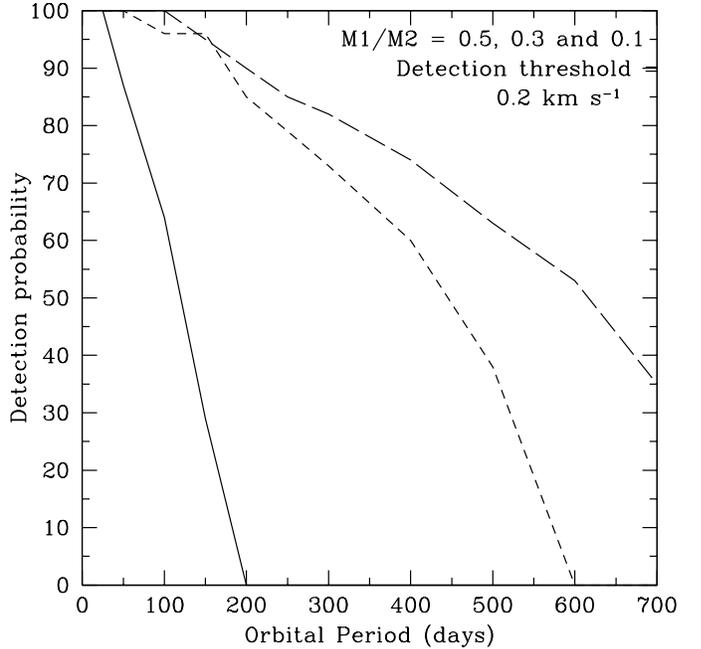}
\caption{\label {simubinaires} Detection probability of massive
  stellar companions as a function of the orbital period for stellar
  companion masses of 0.1, 0.3 and 0.5 M$_{\odot}$ and a solar-mass
  parent star and for a detection threshold of 0.2 \kms. The solid
  line represents the M$_1$/M$_2$ $=$ 0.1 M$_{\odot}$ case, the
  short-dashed line stands for the M$_1$/M$_2$ $=$ 0.3 M$_{\odot}$
  case and the long-dashed line stands for the M$_1$/M$_2$ $=$ 0.5
  M$_{\odot}$ case.}
\end{center}
\end{figure}

We considered a detection threshold of 0.2 \kms\ and calculated the
detection probabilities for 3 different stellar companion masses
(M$_2$/M$_1$ $=$ 0.5, 0.3 and 0.1 M$_{\odot}$). Using the $\sigma$
curve we determined that such a detection threshold corresponds to a
photon-noise uncertainty of 90 \ms\ which covers about 75\%\ of our
stellar sample when considered as an upper limit.  The simulation
results are presented in Fig.~\ref{simubinaires}. These simulations
clearly show that we can identify binary stars with periods longer
than 100 days in our observational configuration.\\

Among the 701 stars with detected CCF peaks we identified 50 binary
stars. Assuming that 50\%\ of the stars are binaries and that 13\%\ of
this population have an orbital period less than 100 days
\citep{DuquennoyMayor91}, we expected to detect 46 of these
systems. This is in good agreement with our results. We demonstrated
above that the detection of systems with an orbital period of up to 1000
days is possible.  Following the statistics from
\cite{DuquennoyMayor91}, we would expect to detect a sample of 90
binary stars with an orbital period less than 1000 days. This suggests
that some of the identified binary stars may have an orbital period
longer than 100 days. The detection ability indicates indeed that we
cover a small part of the systems with an orbital period from 100 to
1000 days. As suggested in Sect.~\ref{candidates} some of the radial
velocity variations of our candidates suggest longer period binary
systems.  The number of RV measurements indeed limits the efficiency
of the detection and a precise determination of the period.\\

\subsubsection{For low-mass companions}
\begin{figure}
\begin{center}
\includegraphics[scale=0.45]{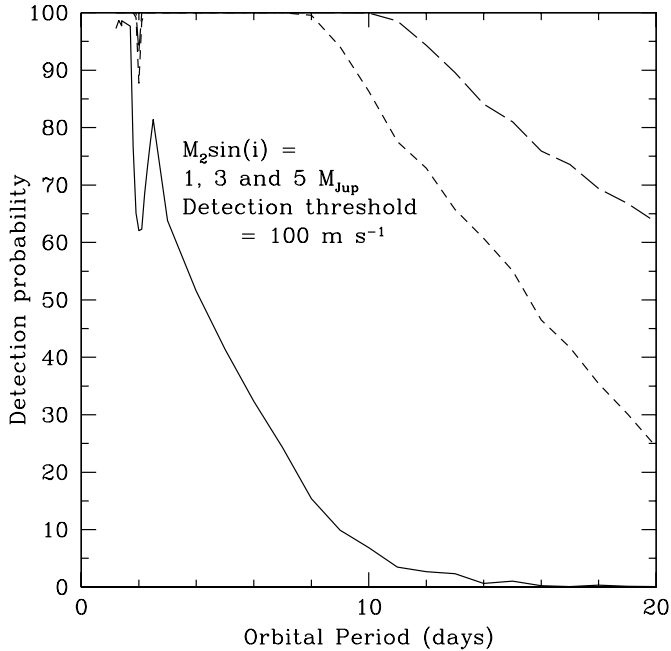}
\caption{\label {detecplan} Detection probability of exoplanets as a
  function of the orbital period for 4 different companion masses
  (M$_1$/M$_2$ $=$ 1, 3 and 5 M$_{Jup}$), a solar-mass parent star
  and a fixed detection threshold of 100 {\ms}. The solid, short-dashed
  and long-dashed lines represent the different
  probabilities for companion masses of 1, 3 and 5 M$_{Jup}$. Note
  that the drop in the detection probability of a 1 M$_{Jup}$
  companion around 2 days decreases due to a stroboscopic
    effect of our exact simulated sampling of one day.}
\end{center}
\end{figure}

For sub-stellar companions, we calculated the detection probabilities
for 3 different companion masses (M$_2$ $=$ 1, 3 and 5 M$_{Jup}$) and
a detection threshold of 100 \ms. Using the $\sigma$ curve we
  determined that such a detection threshold corresponds to a
  photon-noise uncertainty of 37 \ms\/ on the x-axis. Notice that only
  25\% of our sample of stars actually match this lower limit
  value. We computed the detection probabilities for objects with an
orbital period ranging from 1.2 to 20 days. Our simulations show in
Fig.~\ref{detecplan} that we have a 100\% probability of detecting
companions of 1, 3 and 5 \MJ\ with a period less than 2, 9 and 12 days
respectively. The simulations also show that the probabilities of
  detecting companions with less than a Jupiter mass are weak and rapidly
  decrease with the orbital period, with such a detection threshold
  and a small number of RV measurements. However this suggests that
  close-in massive and super-massive giant planets ($>$ 3 \MJ\/) could
  be detected.

As detailed in \citet{Butler06}, 1.2\% of solar-type stars host at
least one hot-Jupiter with a period less than 10 days. Considering our
sample of 584 targets with a distinct peak in their CCF and 5
measurements, we should expect to detect about 7 hot Jupiters. The
statistics of known exoplanets suggest that 33\% of them have a mass
greater than 1~\MJ. With only 25\%\ of our sample (146
  stars), with our observational strategy, our
  survey is expected to detect one massive exoplanet.

\subsection{Extended ability for exoplanet search with the \giraffe\ spectrograph}
We have shown that the \giraffe\ multi-fiber spectrograph could be
used to carry out large exoplanet surveys. Indeed, with a few RV
measurements a systematic error floor of 30 {\ms} could be achieved,
allowing one to explore the massive and the hot-Jupiter exoplanet
populations. With our observational strategy, only 25\%\ of the
observed stars have a photon-noise uncertainty less than the one
obtained for a detection threshold of 100 {\ms}. However this limitation
might be overcome. For example, increasing the number of RV
consecutive measurements from 5 to 10 could reduce the detection
threshold to 1.7 $\sigma$. At small photon-noise uncertainties the
detection threshold is equal to 50 {\ms}. In that case it would be
possible to detect a Jupiter-mass companion with an orbital period up
to 12 days.  Another solution would be to multiply the exposure time
of such faint stars by a factor of two. As a result, the photon-noise
uncertainty would be divided by $\sqrt{2}$, allowing one to widen the
detection threshold of 100~{\ms} to 50\% of the sample. In a similar
way, observing a sample of stars brighter by about one magnitude would
lead to a similar result.  Finally, a multi-fiber facility with
systematic errors of 20~{\ms} would allow us to decrease the detection
threshold to 35~\ms\ at small photon-noise uncertainties, giving
access to the 0.5 \MJ\ exoplanets with short periods. With 10 RV
measurements, objects with periods of up to 8 days would be
detected. Such an efficiency would allow for the discovey of up to 5
new hot-Jupiters in only 10 half-nights.

\section{Conclusion}
A multi-object RV survey enables a very good investigation of a large
sample of stars by distinguishing binaries and low-mass companion
candidates. With only 5 consecutive RV measurements, with
the \giraffe\/ instrument we achieved a systematic RV error of 30 {\ms} on 584
stars. This RV precision reached is 5 times less than announced by the
ESO. On a sample of 816 stars we identified 50 binary systems, 14
exoplanet and brown dwarf candidates, 9 active stars or blended binary
stars and 5 unsolved cases. Given the existence of the BD desert, most
of our candidates might be long-period binary
stars. Assuming the statistics on exoplanets
(\cite{Butler06};Schneider 2007, url: {\it exoplanet.eu}) we could
expect one confirmed exoplanet. However, with only 5 RV measurements,
we could not derive the exact orbital parameters of our candidates and
further higher precision measurements are needed. With this
strategy we reduced by a factor of about 50 a large sample of stars to
a much smaller sample with interesting RV variations for the search of
massive hot Jupiter in only 5 consecutive half-nights. This study
demonstrates that this approach (with optimized data reduction) is
very efficient in searching for massive exoplanets and brown dwarfs
and could bring a key contribution to large and very large surveys. A
similar program carried out with a mono-fiber instrument would
require at least one complete year of observations.\\

The space mission \corot\ was launched successfully on December
2006. This multi-fiber approach could be very useful and efficient in the
follow-up of the large sample of stars that will be observed by the
CoRoT satellite. Indeed a precision of 30 \ms\ obtained with only a
few measurements will allow us to carry out radial velocity
observations over a large sample of stars. Such observations are
necessary to further discriminate binaries, low-mass candidates and
the hot-Jupiter population. In order to characterize the hot-Jupiter
population, the previous knowledge of the two orbital parameters P and
T$_0$ derived from previous high precision photometric observations
will significantly reduce the detection threshold and allow us to
determine the parameters K and V$_0$ at the 1 $\sigma$ level. The
selection made with such a multi-fiber instrument leads to a
significant reduction of the on-telescope time. The most interesting candidates
could thus be observed with more accurate instruments in a second step.\\

New measurements are now needed to confirm the nature of our
substellar companions and derive accurate orbital parameters. To that
purpose we will use the new spectrograph SOPHIE \citep{Bouchy06},
mounted on the 193-cm telescope at the Observatoire de Haute Provence.

\acknowledgements{N.C.S. would like to thank the Fundao para a
  Cincia e a Tecnologia, Portugal, for the grant (reference
  POCI/CTE-AST/56453/2004). This work was supported in part by the EC's
  FP6 and by FCT (with POCI2010 and FEDER funds), within the HELAS
  international collaboration.}

\bibliographystyle{aa} 
\bibliography{biblio} 

\end{document}